\documentclass[a4paper]{spicawStyle}
\usepackage{graphicx,natbib}

\begin{document}

\title{PAHs with SPICA}

\author{O. Bern\'e\inst{1} \and C. Joblin\inst{2,3} \and 
  G. Mulas\inst{4} \and A. G. G. M. Tielens\inst{5} \and J. R. Goicoechea\inst{1}  } 

\institute{
Centro de Astrobiolog\'ia (CSIC/INTA), Madrid, Spain
\and 
 Universit\'e de Toulouse ; UPS ; CESR ; 9 ave colonel Roche, F-31028 Toulouse cedex 9, France
\and
CNRS; UMR 5187; 31028 Toulouse, France
\and
INAF Cagliari, Italy
\and
Leiden Observatory, University of Leiden, Leiden, The Netherlands
 }

\maketitle 

\begin{abstract}

Thanks to high sensitivity and angular resolution and broad spectral coverage, 
SPICA will offer a unique opportunity to better characterize the nature of polycyclic 
aromatic hydrocarbons (PAHs) and very small grains (VSGs), to better use them as 
probes of astrophysical environments. The angular resolution will enable to probe 
the chemical frontiers in the evolution process from  VSGs to 
neutral PAHs, to ionized PAHs and to "Grand-PAHs" in photodissotiation regions 
and HII regions, as a function of G$_0$/n (UV radiation field / density). High sensitivity 
will favor the detection of the far-IR skeletal emission bands of PAHs, which provide 
specific fingerprints and could lead to the identification of individual PAHs. This 
overall characterization will allow to use PAH and VSG populations as tracers of 
physical conditions in spatially resolved protoplanetary disks and nearby galaxies 
(using mid-IR instruments), and in high redshift galaxies (using the far-IR instrument), 
thanks to the broad spectral coverage SPICA provides. Based on our previous 
experience with ISO and Spitzer we discuss how these goals can be reached.
\end{abstract}

\section{Introduction}

The ubiquitous mid-IR emission bands, widely observed in the spectra
of dusty astrophysical sources (from protoplanetary disks to starburst
galaxies), are attributed to the emission of a family of carbonaceous
macromolecules: the polycyclic aromatic hydrocarbons (PAHs). However,
because these bands are due to nearest neighbor vibrations of the C-C
or C-H bonds, they are not specific to individual PAH species.
Therefore,  in spite of their major relevance for astrophysics (as
tracers of the presence of UV radiation fields or star forming
regions in a broader extragalactic context), the identification 
 of a given  PAH molecule in space has yet not been
possible. This contribution explores the new possibilities that could be offered by
SPICA spectrometers to better characterize PAHs:  In the mid-IR, high angular resolution 
will enable to better establish the link between the chemical evolution
of PAHs and VSGs in connection with the evolution of physical
conditions and the formation/excitation of H$_2$. This will then allow to use them as tracers of physical conditions
at low and high redshifts. In the far-IR, we expect to possibly detect the
low-energy vibrational modes of PAHs which are much more
connected to the structure of each molecule and can thus provide
an unprecedented characterization of their nature and evolution 
in space.

\section{Mid-infrared observations of PAH bands}

\subsection{PAHs/VSGs: tracers of physical conditions}\label{trac}

The mid-IR emission of PAHs and VSGs has been well characterized by 
ISO (see e.g. \citealt{pee02,rap05}) and more recently by
Spitzer (see e.g. \citealt{wer04,ber07}). The observed spectrum
usually consists in a set of bands that are most prominent at
3.3, 6.2, 7.7, 8.6, 11.3, and 12.7 $\mu$m. It was established that 
the modification of the shape of this spectrum can be attributed to 
alteration of the chemical structure of the emitting component (\citealt{pee02, hon01})
and that this chemical evolution is strongly connected to the local
physical conditions. In particular, models (\citealt{tie05}) and observations
(\citealt{job96, gal08}) have shown that the variations of the 6.2 (or 8.6) to 11.3 $\mu$m 
band intensity ratio ($I_{6.2}/I_{11.3}$) evolves with the ''ionization parameter"  $\gamma=G_0\times \sqrt{T}/n_H$
where $G_0$ is the intensity of the UV radiation field in Habing's units, $T$ is the
gas temperature and $n_H$ the total hydrogen nuclei density. Following this
work \citet{ber09b} have shown that the combination of the measurement 
of $I_{6.2}/I_{11.3}$ and of the ratio between the H$_2$ 0-0 S(3) and S(2) 
line intensities, respectively at 9.7 and 12.3 $\mu$m, 
allows to derive the individual values of $T$, $G_0$ and $n_H$ when they 
fall in the ranges $T = 250-1500 $K, $n_H=10^4-10^6$cm$^{-3}$, 
$G_0=10^3-10^5$ respectively. 
   
 \begin{figure*}[ht]
  \begin{center}
  \vspace{-2.5cm}

    \includegraphics[width=17 cm]{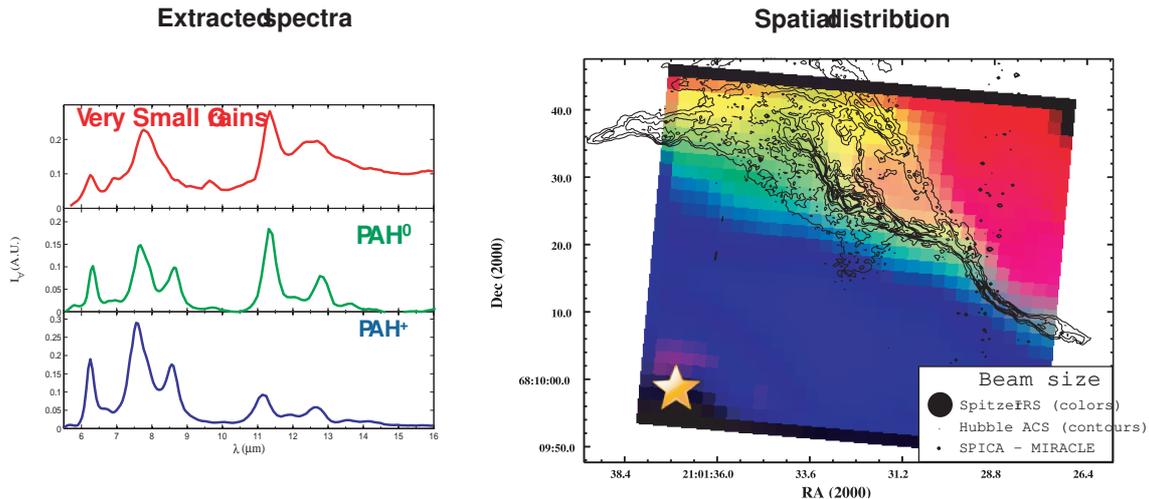}
    \vspace{-3.5cm}

  \end{center}
  \caption{\emph{Left}: Extracted spectra of VSGs, PAH$^0$ and PAH$^+$ in NGC 7023 N Spitzer cube. \emph{Right:} Associated distribution maps of the three populations 
: VSGs in red, PAH$^0$ in green and PAH$^+$ in blue. Colors combine as in an RGB image i.e. green (PAH$^0$)+red (VSGs)=yellow. In contours are shown the filaments
detected in ERE with Hubble.}  
\label{authorf_fig:fig1}
\end{figure*}

\subsection{PAHs/VSGs: role in the formation of H$_2$ (?)}
  
 H$_2$ is the most abundant molecule in the universe but its formation mechanism is still an open question.
 It is however clear that H$_2$ forms at the surface of grains, as the gas-phase routes 
 are too inefficient under standard ISM conditions (\citealt{gou63}). The formation rate of H$_2$ in photodissociation regions (PDRs) was found to be 
 larger than that derived from the ÓclassicalÓ formation mechanism at the surface of cold grains 
(\citealt{hab04}). Furthermore, \citet{job00} have shown that PAH/VSG and H$_2$ emissions
 spatially correlate. In this context, it has been considered that PAHs and/or VSGs could play
 a role in the catalysis of H$_2$ formation. 

\subsection{PAH and VSG chemistry with SPICA}

The study of galactic PDRs, where one can
resolve the variations of PAH and VSG emissions as the UV field is attenuated,
is crucial for the understanding of their chemical evolution and their link 
with H$_2$ formation. Until now, PDRs have been extensively observed in the 
mid-IR but only at low angular resolution. There is nevertheless strong evidence that 
the zone where ``everything happens" is in fact very thin and not resolved by Spitzer 
and ISO. Indeed, the evolution from VSGs to PAHs occurs in a region of less than 2 
magnitudes where physical conditions vary significantly. Considering that the molecular 
cloud has a density of 10$^{5}$ cm$^{-3}$ this extinction represents, at a distance of 500 
pc, an angular scale of $\sim$1.3" requiring subarcsecond resolution to be probed. 
Observational evidences of this sharp variation of density and radiation field are numerous: 
\citet{ber08} have shown that the transition from VSGs to PAHs, traced by the Extended 
Red Emission, is very sharp (filaments of $\sim$1") in NGC 7023 (Fig. 1). The same 
conclusions are found for NGC 2023 (Pilleri et al. in prep.). H$_2$ 2.12 $\mu$m
high angular resolution data also evidence this very thin transition (\citealt{lem99}). 
Radio interferometric (\citealt{ger09}), and near-IR H$_2$ (\citealt{hab05}), 
observations have evidenced the arcsecond scale chemical stratification in the Horshead 
PDR. In the same region, a steep density gradient has been put forward by \citet{pet05} 
as well as a possible destruction of PAH molecules occurring within these small spatial 
scales. In Monoceros R2 and the Orion bar, H$_2$  2.12 $\mu$m observations also suggest 
the presence of a thin membrane separating the molecular gas from the HII region 
(Walmsley et al. 2000). Unfortunately, it has for now been impossible to spatially
resolve this frontier with spectral maps in the mid-IR as the best achieved angular resolution
with Spitzer IRS was 3.6". This prevents from understanding the link between the chemical 
evolution of PAHs and VSGs and the origin of ERE or H$_2$ formation processes.
The MIRACLE camera onboard SPICA will be particularly suited to solve this
observational issue. With a subarcsecond resolution, while having a low spectral
resolution adapted to PAH bands, MIRACLE will enable, for the first time, to probe
these chemical frontiers that are crucial for our understanding of PAHs/VSGs and H$_2$
photochemistry (Fig. 1). One key point is the use of the imaging with band
filters at R$\sim$5. The fitting of the limited number of spectral points  with a linear 
combination of PAH$^0$, PAH$^+$ and VSG spectra (see templates adapted from Fig. 1
in  \citealt{ber09}) will provide, for each position in the images,
a good estimation of the full mid-IR spectra as well as the spatial distribution 
of PAH and VSG populations, {\bf without spending any time doing spectroscopy}.
This can only be achieved if the number of filters in the 5-14 $\mu$m range is sufficient ($>$10).
Furthermore, the large field of view (FoV) provided by SPICA in the mid-IR will allow to map
much larger regions at once. For instance, the NGC 7023 reflection nebula fits entirely
in the MIRACLE FoV (i.e. North, East and South PDRs at the same time).

\section{PAHs in protoplanetary disks}

 \begin{figure}[ht]
  \includegraphics[width=9cm]{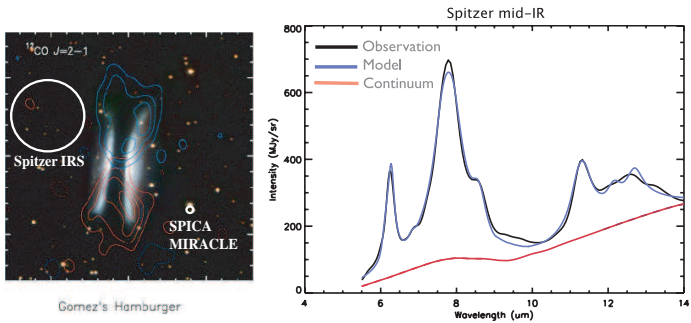}
  \caption{\emph{Left:} Hubble Space telescope observations of Gomez's Hamburger in the visible.  Overlaid in colors the $^{12}$CO J =2-1 emissions 
  for velocities between 0.9 and 4.1 km s$^{-1}$, Doppler shifted towards blue (north) and red (south) from \citet{buj09}. White
  circles are the beams of Spitzer IRS and SPICA. \emph{Right:} The Spitzer-InfraRed Spectrograph spectrum of Gomez's Hamburger in 
  black and adjusted PAH/VSG model (from Bern\'e et al 2009) in blue.}  
\label{fig3}
\end{figure}

It has become clear, in the recent years, that the disks of gas and dust that form around young stellar objects are the cradles
of planetary formation (see e.g. recent observations of exoplanet embedded in the disk of Formalhaut by \citealt{kal08}). 
PAHs and VSGs are know to be present in the gas-rich protoplanetary disks and play multiple roles: 
\emph{(i)} they can be considered as building blocks from which larger bodies can form by aggregation, \emph{(ii)}
because they are a major source of optical thickness in the UV (\citealt{dra07}) they shield the gas from photodissociation
(\citealt{dul07, ber09}) and slow the photo-evaporation process (\citealt{ale08}).
Only very few studies from the ground have enabled to look at the properties of PAHs within protoplanetary
disk, and only in imaging through broad-band filters (\citealt{lag06, gee07}). MIRACLE will spatially resolve such disks (see e.g. Fig. 3), and
with imaging at R$\sim$5 provide the maps of PAHs$^{0/+}$ and VSGs. Using the 
methods described in Sect.\ref{trac} we will be able to obtain the physical properties of different regions of the
surface at the disk.

\section{Redshifted mid-IR PAH bands}


Spitzer has brought clear evidence that PAH bands are present
in the emission of galaxies dominated by star formation at $z>2$ (see e.g. \citealt{pop08}). 
Star formation rates (SFR) of galaxies can then be estimated by relating the
PAH luminosity to the total IR luminosity (see e.g. \citealt{bra06}) and then
the total IR luminosity to the SFR using the \citet{ken98} law.
Recent studies have also intended to relate the ionization fraction / size
of PAHs to the SFR (\citealt{odo09}). SPICA will enable
to better characterize the relationship between SFR and PAH emission,
and will provide PAH spectra of galaxies at higher red-shifts (see cosmological
implications in the extragalactic section of these proceedings). 
One important point is that with it's unprecedented sensitivity, SPICA will allow
to observe {\bf both the PAH and H$_2$ emissions} at high $z$. 
As show by \citet{ber09b} this can be very useful to learn about the 
physical conditions prevailing in the emitting environment.

\section{Far-IR modes of PAHs}

 \begin{figure*}[ht]
  \begin{center}
    \vspace{-2cm}

    \includegraphics[width=9 cm]{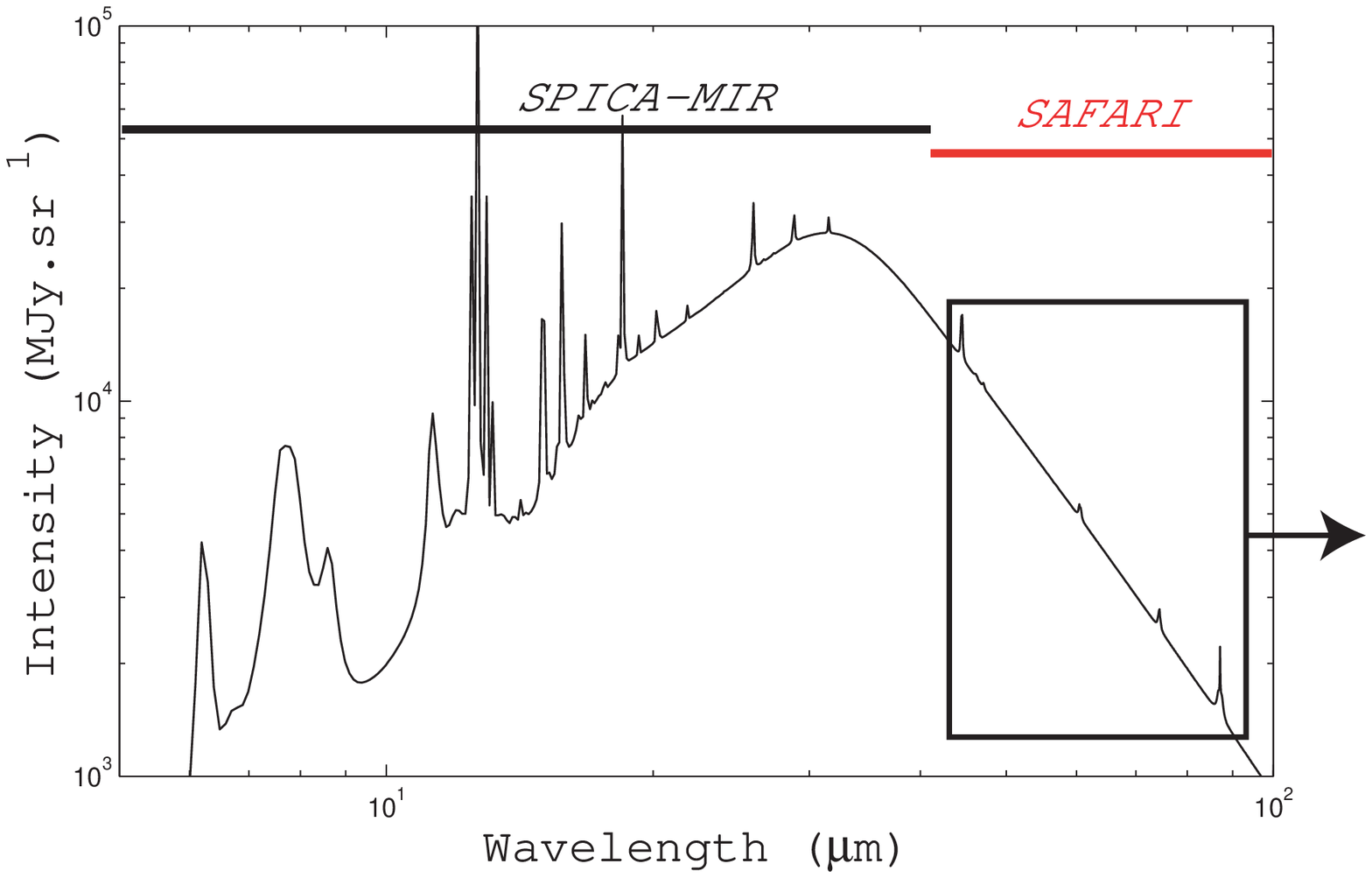}
        \includegraphics[width=7 cm]{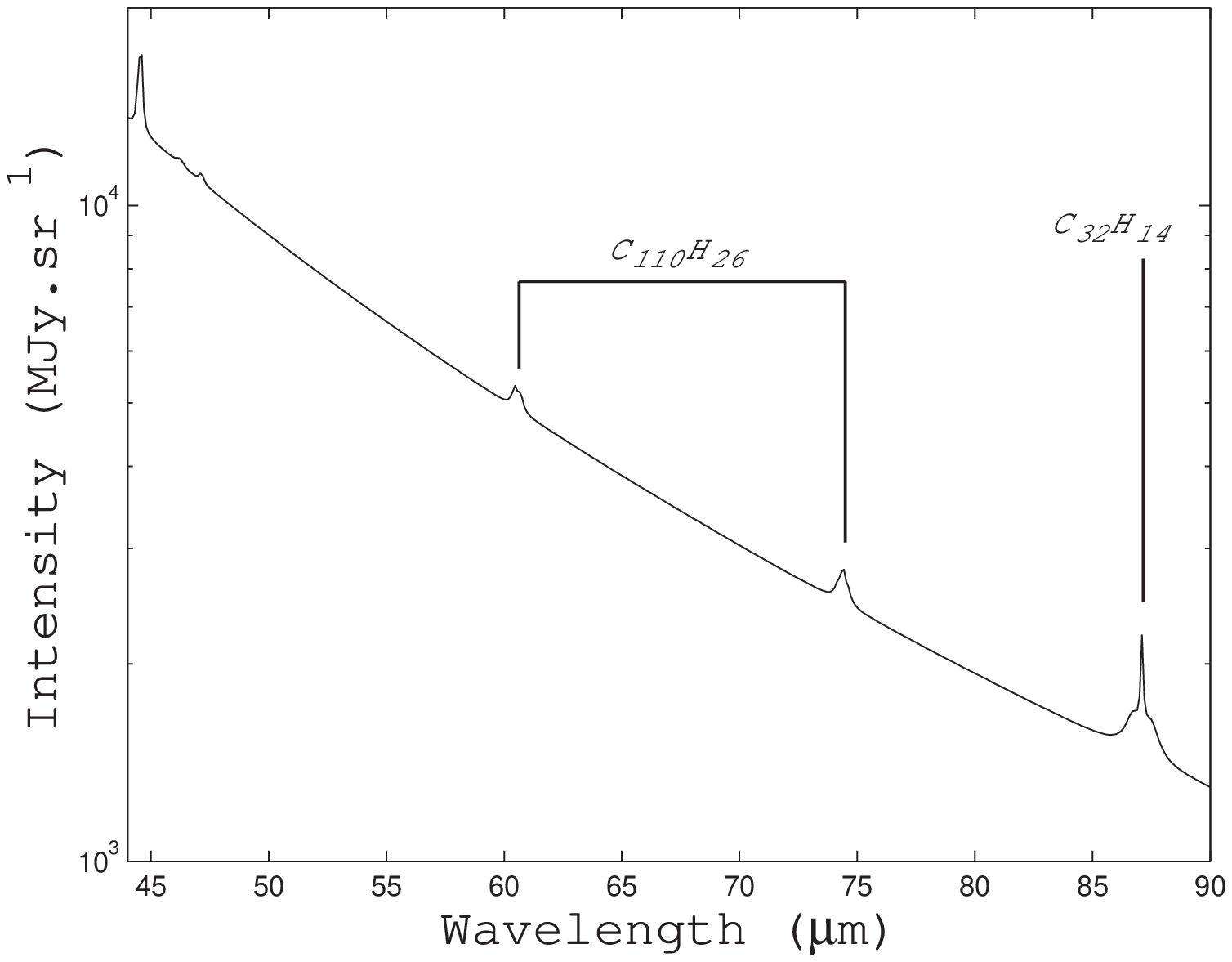}

  \end{center}
  \caption{Modeled spectrum of a PDR with a $G_0=10^5$ radiation field, column density of N$_H=10^{20}$cm$^{-2}$ and 20\% of carbon locked in PAHs and assuming
  that the mid-IR bands are due to C$_{110}$H$_{23}$ and C$_{32}$H$_{14}$.
  The adopted spectral resolution is $\sim$ 1000. The mid-IR PAH bands in the spectrum and the continuum emission calculated with the model 
  of \citet{dra07}. The far-IR emission PAH bands are calculated using the photochemical model of \citet{mul06} using IR cross 
  sections calculated by \citet{mal07} and I. Cami (private com.). }  
\label{authorf_fig:fig1}
\end{figure*}

Far-IR emission bands of PAHs would specific fingerprints of
individual molecules. These low-energy vibrations  involve the
bending of the whole PAH skeleton (mostly out-of-plane) and are thus
intrinsically related to the structure of each possible PAH carrier.
Unfortunately, PAHs tend to release their vibrational energy mostly
through mid-IR emission, and thus their far-IR bands are expected to
be very weak.  As an example, ~0.2\% of the total UV energy absorbed by
a PAH like coronene,  will be emitted in  FIR band emission  (\citealt{job02}).  
Therefore, the PAH  FIR band emission is expected to be difficult to
detect in space. Nonetheless, it can be shown that if all the mid-IR
emission observed in the ISM was due to only a few large PAH
molecules, their far-IR emission band should be detectable even with a low
($<$100) signal-to-noise ratio (see \citealt{mul06} and Fig. 3).  
ISO/LWS detected several unidentified far-IR bands (\citealt{cer02, goi04})
 although none could specifically be attributed to PAH emission. The non detection of such bands with ISO
combined with other evidence in the mid-IR (\citealt{pee02}), suggest
that a scheme in which there are only a few different PAH molecules 
is unlikely. If instead, there are 50 different PAHs
responsible for the mid-IR band spectrum, the strongest
band emitted by all of them in the far-IR will have a peak intensity of $\sim$1\% of the
thermal continuum intensity. While this emission may
seem extremely weak, it is not  impossible to detect it, given the
progresses made in far-IR space instruments, and the prospects of much
improved sensitivity with the SPICA/SAFARI spectrometer. If there were
really 50 different PAHs in space,  the problem might come from
somewhere else: the relatively uncertain wavelength position of these
PAH bands, and from spectral confusion. Indeed, the more numerous
interstellar PAHs are,  the more numerous far-IR PAH bands there will
be. Fortunately, recent progress in the field of interstellar PAHs
suggests that there are in fact only a limited number of large and 
compact PAHs in space (see \citealt{tie08} for review), that 
can resist the harsh interstellar conditions thanks to
their ability to redistribute the energy they absorb efficiently
inside the molecule, these are the so called ÒGrand-PAHsÓ. The
instantaneous broad band coverage of SAFARI-FTS will be specially
adapted for deep searches of far-IR bands where the wavelength
positions are not constrained spectroscopically and can appear in the
whole domain.
The discovery of the specific  PAH carriers responsible of the
widespread mid-IR PAH emission (local and extragalactic) will
constitute a tremendous step forward for our understanding of the
chemical complexity of the universe. Before that, unprecedented
efforts in parallel  instrumental, theoretical, laboratory and
observational aspects will have to be carried out. Herschel will be
limited in terms of wavelength coverage and sensitivity in the far-IR,
and thus SPICA/SAFARI can represent our first chance to detect
specific PAHs.

\begin{acknowledgements}

OB is supported by JAE-Doc CSIC fellowship.
JRG was supported by a Ramon y Cajal research contract from the spanish
MICINN and co-financed by the European Social Fund.
OB, and CJ acknowledge the french national program PCMI.

\end{acknowledgements}

\end{document}